\documentclass[aps,pra,reprint,showpacs,nofootinbib,superscriptaddress]{revtex4-1}
\usepackage{graphicx, endnotes, appendix}
\usepackage{epsfig}
\usepackage{bm}
\usepackage{slashbox}
\usepackage{color}
\usepackage{float}
\usepackage{dcolumn}
\usepackage{multirow} 
\usepackage{hyperref}
\usepackage{amsmath}
\usepackage{amsfonts}
\usepackage{mathtools}
\graphicspath{{.}}
\usepackage{bm, listings, verbatim,graphicx, amsfonts, dsfont, bbold, xcolor}
\newcommand{\be}{\begin{equation}}
\newcommand{\ee}{\end{equation}}
\newcommand{\ba}{\begin{eqnarray}}
\newcommand{\ea}{\end{eqnarray}}

\newcommand{\ket}[1]{| {#1} \rangle}
\newcommand{\expect}[1]{\langle {#1} \rangle}

\begin{document}

\title{Benchmarking Noise Extrapolation with OpenPulse}
\footnotetext{This manuscript has been authored by UT-Battelle, LLC under Contract No. DE-AC05-00OR22725 with the U.S. Department of Energy. The United States Government retains and the publisher, by accepting the article for publication, acknowledges that the United States Government retains a non-exclusive, paid-up, irrevocable, world-wide license to publish or reproduce the published form of this manuscript, or allow others to do so, for United States Government purposes. The Department of Energy will provide public access to these results of federally sponsored research in accordance with the DOE Public Access Plan. (http://energy.gov/downloads/doe-public-access-plan).}

\author{J. W. O. Garmon}
\affiliation{Department of Physics}
\affiliation{Bradley Department of Electrical \& Computer Engineering, Virginia Tech, Blacksburg, VA 24061}
\author{R. C. Pooser}
\affiliation{Computational Sciences and Engineering Division, Oak Ridge National Laboratory, Oak Ridge, TN 37831}
\author{E. F. Dumitrescu}
\affiliation{Computational Sciences and Engineering Division, Oak Ridge National Laboratory, Oak Ridge, TN 37831}
\begin{abstract}
Distilling precise estimates from noisy intermediate scale quantum (NISQ) data has recently attracted considerable attention~\cite{Kandala2018}. In order to augment digital qubit metrics, such as gate fidelity, we discuss analog error mitigability, i.e. the ability to accurately distill precise observable estimates, as a hybrid quantum-classical computing benchmarking task. Specifically, we characterize single qubit error rates on IBM's \textit{Poughkeepsie} superconducting quantum hardware, incorporate control-mediated noise dependence into a generalized rescaling protocol, and analyze how noise characteristics influence Richardson extrapolation-based error mitigation. Our results identify regions in the space of Hamiltonian control fields and circuit-depth which are most amenable to reliable noise extrapolation, as well as shedding light on how low-level hardware characterization can be used as a predictive tool for uncertainty quantification in error mitigated NISQ computations. 
\end{abstract}
\maketitle

\section{Introduction}

Quantum information processing technologies promise to offer dramatic speedups for a variety of computational tasks \cite{NielsonMichaelA.andChuang2001}. It has, however, been clear for some time that environmental interactions and control errors pose a significant barrier to the coherent operation of a quantum computer. While quantum error correction~\cite{Shor1995} is the de-facto (i.e. only known scalable) solution to suppressing the effects of information corrupting processes, error correction schemes are infeasible on present NISQ devices due to the overwhelming qubit overhead, lack of feedforward control, and gate fidelity requirements. Due to the growing availability and the exciting prospects of programmable NISQ information processing devices, these constraints have forced researchers to consider new methods for improving computational results. A variety of alternative schemes, broadly referred to as quantum error mitigation (QEM), have recently been introduced~\cite{Temme2017,McClean2017,Otten2019,McArdle2019}, with some having been demonstrated at small scales \cite{Dumitrescu2018, Kandala2018, Klco2018, Colless2018}. 

Recent QEM progress leads to the natural question of how to determine fundamental scalability and precision limitations of each method. While tomographic mappings~\cite{DAriano2003, Mohseni2008, NielsonMichaelA.andChuang2001, Blume-Kohout2017} and randomized benchmarking~\cite{Knill2008} serve as good digital gate measures, with well understood scalings, similar metrics for defining the limits and precision of hybrid quantum-classical computations using QEM as a central component are underdeveloped. We therefore introduce mitigability as a hybrid quantum-classical benchmark which will provide insight into the fundamental limits of NISQ algorithm performance. 

While mitigability can be defined and characterized for any QEM strategy, this work focuses on error elimination by extrapolation at the analog-level by stretching microwave gate pulses~\cite{Kandala2018}. Analog control theory is central to the development of high fidelity quantum operations ~\cite{Mabuchi2005,Wiseman2009,Brif2010} but in our approach analog control systems are used to {\em amplify} quantum noise. The amplified noise levels can then be used both to characterize quantum hardware and as input for data post-processing QEM schemes. In this regard, QEM mitigability serves as a hybrid benchmark in the analog domain, with direct implications on the performance of quantum processors in the hybrid quantum-classical computational paradigm.

The remainder of our work is organized as follows. We first review the method of extrapolation to the noiseless limit, its implementation by pulse-stretching, and define mitigability based on an estimator error. We then perform an experimental analysis on a representative set of analog, single qubit operations implemented on the IBM \textit{Poughkeepsie} device using the OpenPulse control framework~\cite{McKay2018}. Lastly we analyze our results and discuss the future research directions necessary to more accurately predict the performance of error mitigated NISQ computations. 

\section{Error removal by extrapolation}
\label{sec:formalism}
Consider a quantum circuit $U_I$ unitarily evolving an initial state $\rho_0$ to an ideal logical state $\rho_I$. In the presence of unitary and stochastic errors, the final state $\rho (\bm{\epsilon})$ differs from $\rho_I$ and can be expressed as a function of the noise parameters $\bm{\epsilon} = (\epsilon_1,...,\epsilon_n)$. In order to simplify analysis let us assume all error sources may be represented by a single effective noise parameter $\epsilon$. Richardson's deferred approach to the limit~\cite{Richardson1927} (i.e. noiseless extrapolation) has recently been proposed~\cite{Temme2017} and implemented~\cite{Dumitrescu2018, Klco2018, Kandala2018} as a NISQ-era QEM technique to construct a noiseless estimator for an observable $\expect{A}$ given a set of states $\{\rho(\epsilon_i)\}$ with effective error rates $\{\epsilon_i\}$. 

To see how Richardson extrapolation reduces the effects of noise, we follow Ref.~\citenum{Temme2017} and assume that $\rho(\epsilon)$ can be Taylor-expanded about the ideal logical state $\rho_I$ in powers of $\epsilon$ when $\epsilon \ll 1$. The expansion reads
\begin{equation}
\rho(\epsilon) = \sum_{k=0}^\infty  \frac{\partial^k_\epsilon \rho(\epsilon)|_{\epsilon=0}}{k!} \epsilon^k = \rho_I + \sum_{k=1}^\infty \rho^{(k)}(0) \epsilon^k
\end{equation}
where $\rho^{(k)} = \frac{\partial^k_\epsilon \rho(\epsilon)|_{\epsilon=0}}{k!}$ represents the matrix coefficients given by the $k^\text{th}$-order derivative with respect to $\epsilon$ and $\rho^{(0)}(\epsilon) = \rho_I$. It follows that expectation values may likewise be expanded as $\expect{A}(\epsilon) = \sum_{k=0}^\infty \expect{A}^{(k)} \epsilon^k$ where $\expect{A}^{(k)} = \text{Tr}[A \rho^{(k)}]$ at each order. 

Extrapolation involves increasing the error rate by n factors $\{a_i\}$, so that $\{\epsilon_i\} = \{a_i \epsilon \}$, and evaluating the corresponding expectation values ${\expect{A}(a_i \epsilon)}$. We may take $a_0=1$ to represent the original data point. For each $a_i$ the expectation value is expanded with respect to the rescaled error rates $a_i \epsilon$, so that $\expect{A}(a_i \epsilon) = \sum_{k=0}^\infty \expect{A}^{(k)} (a_i \epsilon)^k$. Imposing normalization, $\sum_i b_i = 1$, and a set of eliminator constraints, $\sum_i b_i a_i^k = 0 \; \forall i \in (1,n-1)$, one may construct a noiseless estimator through the linear combination 
\begin{eqnarray}
\label{eq:noise_rescaling_expansion}
\expect{A}_{E} &=& \sum_{i=0}^{n-1} b_i {\expect{A}(a_i \epsilon)} \nonumber \\
&=& \expect{A}_I \sum_{i=0}^n b_i + \sum_{k=1}^\infty \expect{A}^{(k)} \epsilon^k \sum_{i=0}^n b_i a_i^k \nonumber \\
&=& \expect{A}_I + \mathcal{O} (\epsilon^n)
\end{eqnarray}
where $\expect{A}_I = \text{Tr}[A \rho_I]$ and $\expect{A}_E$ is the estimated expectation value. If each $a_i$-factor expectation value is obtained independently the estimator variance becomes the weighted sum of the individual variances $\text{Var}[\expect{A}_I^{Est}] = \sum_i b_i^2 \text{Var}[\expect{A}(a_i \epsilon)]$. The estimator error $\Delta_A = |\expect{A}_I - \expect{A}_E|$ is an obvious target metric which quantifies the effectiveness of noiseless extrapolation and the validity of its underlying assumptions.

\subsection{Rabi Stretching}

\begin{figure}[tb!]
    \centering
    \includegraphics[width=\columnwidth]{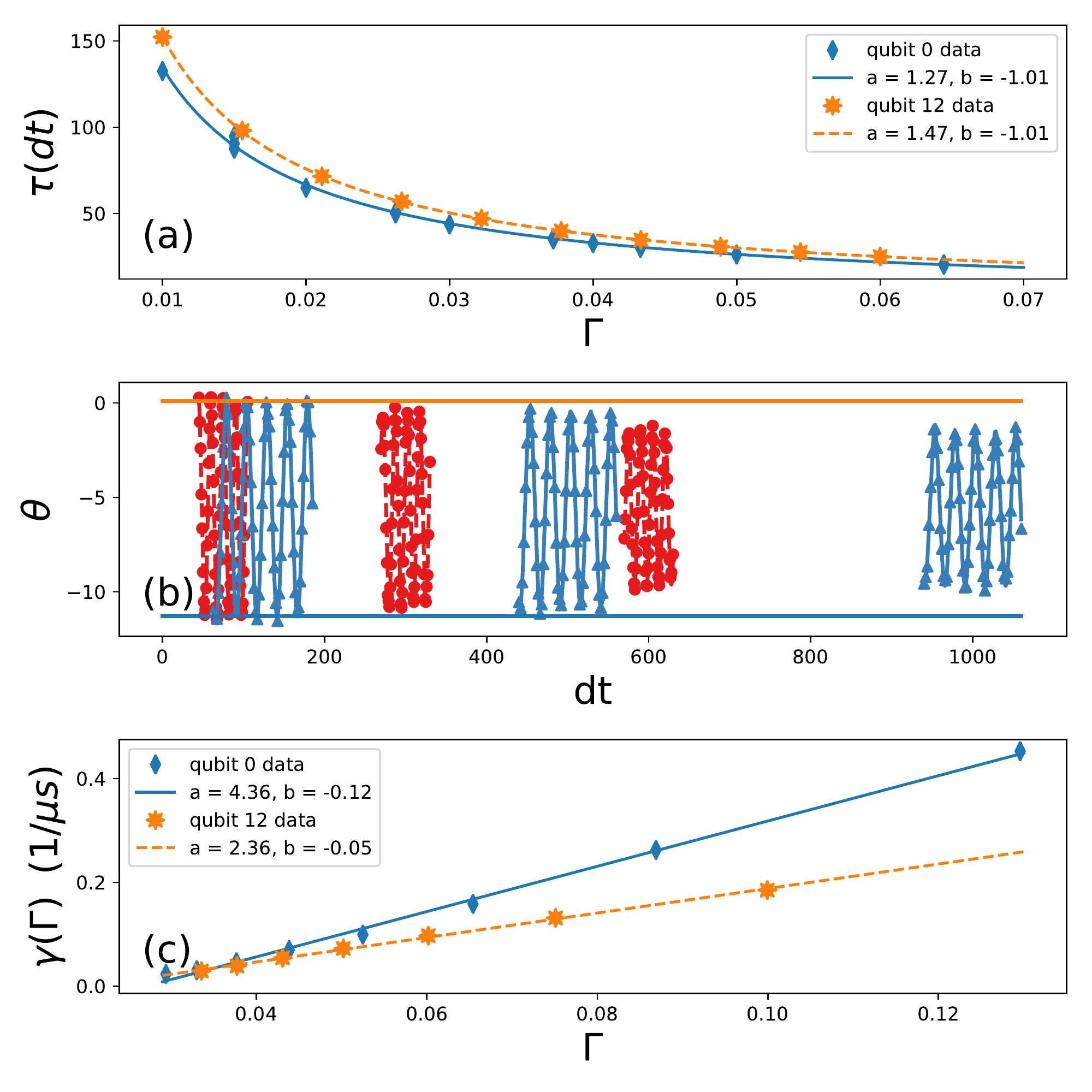}
    \caption{
    (a) Observed Rabi period $\tau$ as a function of the drive pulse amplitude $\Gamma$. The data closely matches the expected $\tau(\Gamma) = a \Gamma^b$ with $b=-1$ inverse power law. The least squares fit is inverted to program Rabi oscillations at various stretch factors (see Sec.~\ref{sec:Results}). 
    (b) Rabi oscillation amplitude decay as a function of time for Rabi Periods of $\tau = 15$ (red circles) and $\tau = 25$ (blue triangles) where our time units are in terms of the OpenPulse default $dt=3.55ns$.
    (c) The amplitude dependent component of the relaxation rate $\gamma(\Gamma) = \tilde{\gamma}(\Gamma) - 1/T_1$ as a function of drive amplitude $\Gamma$. The effective relaxation rate $\tilde{\gamma}(\Gamma)$ is determined by fitting an exponential decay to continuous Rabi flopping at various driving amplitudes. Blue diamonds (orange stars) shows data obtained for qubit 0 (12) and the solid (dashed) lines correspond to fits with coefficients specified in the legend.}
    \label{fig:1}
\end{figure}

Extrapolation based on analog pulse stretching~\cite{Kandala2018}, which goes beyond digital gate repetition methods~\cite{Dumitrescu2018, Klco2018}, has recently been realized. Assume the quantum dynamics are described by a generic master equation $\partial_t{\rho(t)} = -i[H(t),\rho(t)] + \epsilon \mathcal{L}[\rho(t)]$ with the first commutator term representing unitary Hamiltonian evolution and the latter component representing some undesirable noisy dynamics. The authors of Ref.~\citenum{Temme2017} have shown that a set of states $\rho(a_i \epsilon)$ with stretched error rates $a_i=c$ may be generated contingent upon $\epsilon \mathcal{L}$ being invariant under a time rescaling of the form $t\rightarrow t'=t/c$ and independent of Hamiltonian rescaling, i.e. $H(t)\rightarrow H'(t')=H(t/c)/c$. 

Let us restrict ourselves to the minimal model of single qubit operations. Then, working within the qubit's rotating reference frame the Hamiltonian consists of only a single driving term $H = \Gamma \sigma^x$. Next, we consider single qubit logical circuits consisting of $M$ of Rabi oscillations. That is, two evolutions are logically equivalent if both realize $M$ Rabi flops. This circuit could also be interpreted as an $M$ $X$-gate bit-flip circuit in a digital setting. 

We now consider the details of programming Rabi oscillations at desired periodicities. Beginning in $\ket{0}$, the probability of the excited state $\ket{1}$ to be populated at time $t$ is given by the Rabi formula $|c_1(t)|^2 = \frac{\Gamma^2}{\Gamma^2+\Delta^2} \sin^2{\omega t}$ where $\Gamma$ will be the driving amplitude specified in OpenPulse~\cite{McKay2018}, $\Delta=\omega_{12}-\omega_D\approx0$ is the driving-qubit frequency detuning, and the Rabi frequency $\omega = 2\pi/\tau = \sqrt{\Gamma^2+\Delta^2}$ is directly proportional to the driving amplitude under a resonant drive. By fitting the observed Rabi oscillations to a sinusoid for a range of driving amplitudes we are able to determine a quantitative relationship between Rabi period and drive amplitude. The fitted power law $\tau(\Gamma) = a \Gamma^b$ closely matches the theoretically predicted inverse proportionality as illustrated in panel (a) of Fig.~\ref{fig:1}. Due to OpenPulse constraints at the time of writing (i.e. without a digitizing discriminator as in an OpenPulse level 2 measurement~\cite{McKay2018}) we report the magnitude of Rabi oscillations in terms of a state dependent phase difference observed in heterodyne measurement $\theta_{10} = \theta_{\ket{1}} - \theta_{\ket{0}}$~\cite{Krantz2018} as illustrated in Fig.~\ref{fig:1} (b). 

\subsection{Amplitude dependent noise rescaling}

Recent extrapolation demonstrations have assumed that the error rate per unit time is invariant under pulse stretching~\cite{Temme2017, Kandala2018}. In this setting, additional errors accumulate due to a longer evolution and the modified noise factor increases linearly in the temporal stretch factor. In our continuously-driven Rabi flopping experiments we observe that this is generally not the case. This can be seen in Fig~\ref{fig:1} panel (b) where we have plotted $\theta_{10}$ at $M=5,20,$ and $40$ oscillations (from left to right) for Rabi periods of $\tau=15$ red (circles) and $\tau=25$ (blue triangles). By the 40th (rightmost data) cycle the $\tau=15$ evolution has decayed further than the $\tau=25$ experiment despite a significantly more rapid evolution. By continuously driving Rabi oscillations at a fixed amplitude and fitting the decay to an exponential function, we recover an effective $\Gamma$-dependent relaxation rate $\tilde{\gamma} (\Gamma)$. We model the generalized relaxation rate as $\tilde{\gamma} (\Gamma) = 1/T_1 + \gamma(\Gamma)$ and subtract out the bare $T_1$ relaxation rate (determined by exciting the $\ket{1}$ state and allowing it to relax without continuous driving) in order to determine the drive dependence of the noise. The amplitude dependent component, suspected to arise from leakage into the transmon's $\ket{2}$ state, is well described by a linear fit as illustrated in Fig.~\ref{fig:1} panel (c).

Let us now consider a generalized, drive dependent noise expansion parameter. Using $\tau_i = c_i \tau_0$, with $\tau_0$ the bare Rabi periodicity, we integrate the error rate per unit time for duration of the program writing $\epsilon(M, \tau_i(\Gamma), T_1) = \int_0^{M \tau_i } dt \tilde{\gamma}(T_1, \tau_i (\Gamma)) = M \tau_i * (1/T_1 + \gamma(\Gamma))$. Here $\epsilon(M, \tau_i(\Gamma), T_1)$ represents the error accumulated after $M$ Rabi cycles generated by a driving term $\Gamma_i=(\tau_0/ac_i)^b$ and $T_1$ is the bare relaxation time. While we have loosened the assumption of Hamiltonian independence, the time-translation invariant noise assumption remains.

\section{Results}
\label{sec:Results}

We are now in a position to investigate the effectiveness of QEM by pulse stretching on different qubits. To do so, we consider five logical programs which are indexed by and consisting of $M \in (5,20,40,80,160)$ Rabi oscillations. For each $M$ we take the initial noisy program to evolve with a Rabi periodicity of $\tau_0=10 dt$ ($dt=3.55ns$ by default in OpenPulse), a similar timescale to that of a digitized $X_{\pi/2}$ gate~\cite{McKay2018}. 

\begin{figure}[t!]
    \centering
    \includegraphics[width=\columnwidth]{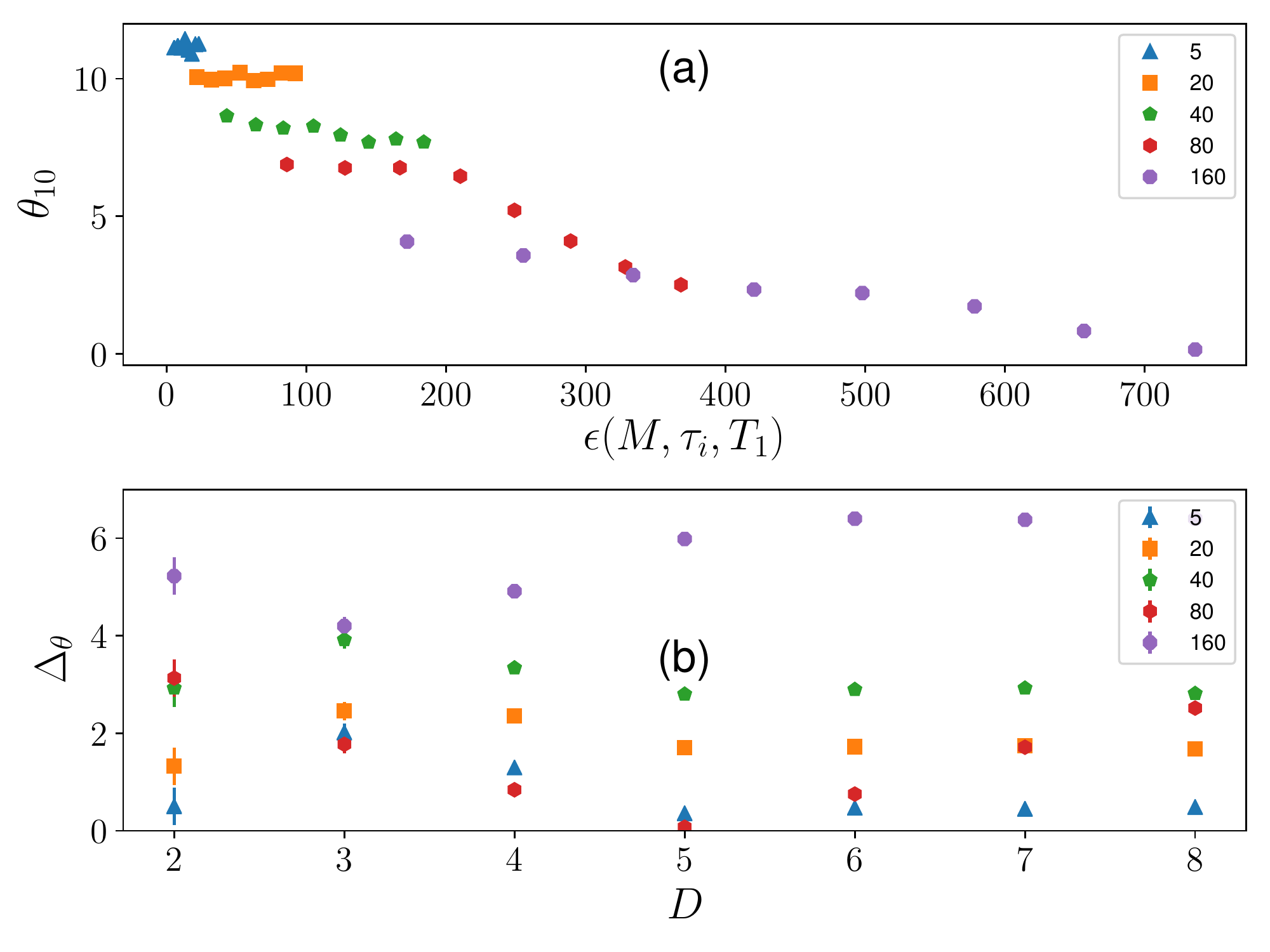}
    \caption{
    (a) Rabi oscillation IQ phase difference $\theta_{10}$ for qubit 0 as a function of the noise factor $\epsilon(M, \tau_i(\Gamma), T_1)$. The time rescaled expectation values are indexed by the number of cycles as indicated by the legend. (b) Estimator error from a linear extrapolation utilizing increasing numbers of data points. Error bars represent the relative increase in the estimator standard deviation. 
    }
    \label{fig:noise_factors_vs_rabi_amplitude_qb0}
\end{figure}

\begin{figure}[t!]
    \centering
    \includegraphics[width=\columnwidth]{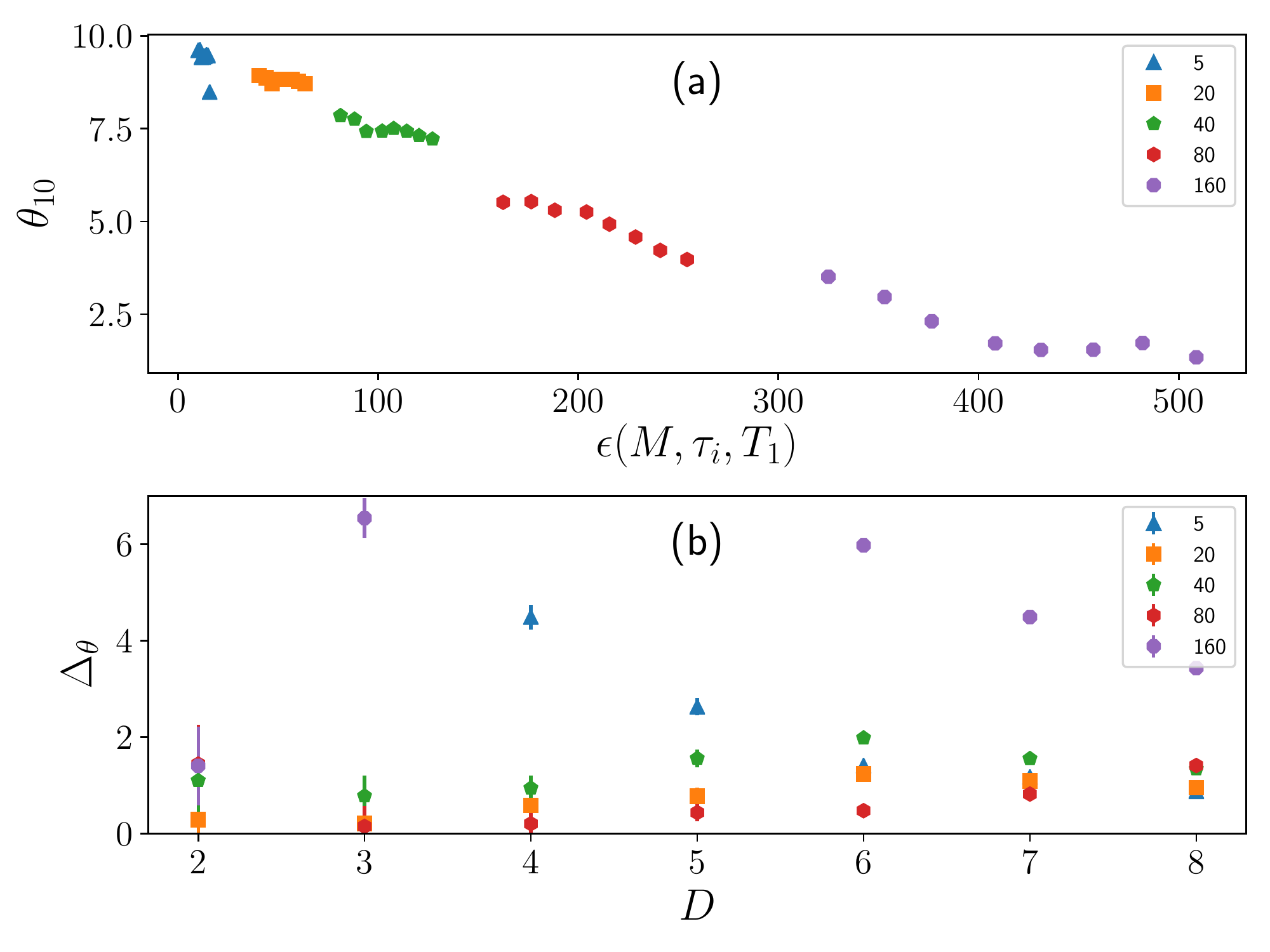}
    \caption{
    Rabi phase difference (a) and estimator error (b) as in Fig.~\ref{fig:noise_factors_vs_rabi_amplitude_qb0} but for qubit 12. Note the increase in linearity and the reduced range for the noise factors.}
    \label{fig:noise_factors_vs_rabi_amplitude_qb12}
\end{figure}

Next, we run additional $M$ Rabi flop experiments in which the Rabi periodicities are stretched by factors of $c_i=(1,1.5,2,2.5,3,3.5,4,4.5)$. The top panels in Figs.~\ref{fig:noise_factors_vs_rabi_amplitude_qb0},~\ref{fig:noise_factors_vs_rabi_amplitude_qb12} plot the Rabi oscillation decay, as indexed by the legend starting with $M=5$ (blue triangles -- top left) to $M=160$ (purple circles -- bottom right), as a function of the computed noise factors. The amplitude dependent noise factors $\epsilon(M, \tau_i(\Gamma), T_1)$ can now be inserted into Eq.~\ref{eq:noise_rescaling_expansion} in order to solve for the ${b_i}$ weights which satisfy the normalization and elimination constraints.

% noise factors
Before implementing the rescaling protocol we briefly examine the Rabi oscillation's dependence on the noise factor $\epsilon(M, \tau_i(\Gamma), T_1)$ as illustrated in the top panels of  Figs.~\ref{fig:noise_factors_vs_rabi_amplitude_qb0},~\ref{fig:noise_factors_vs_rabi_amplitude_qb12}.  First, note that the $M=5,20,40$ cycles approximately linear decay (modulo the outlying $M=5$ data point in Fig~\ref{fig:noise_factors_vs_rabi_amplitude_qb12} (a) for which our automated fit errored) and the $M=80,160$ cycle's non-linear behaviour serves to qualitatively delineate the length of programs for which first and higher order expansions are appropriate. Indeed, the more linear behavior for the $M=80$ data for qubit 12 compared with qubit 0 suggests that its error rate is lower, better controlled, and therefore that qubit 12 is a better choice for QEM by extrapolation.

Regarding the $M=80,160$ cycles data, practical considerations hamper high polynomial order estimates. This is mainly due to the rapid accumulation of statistical errors arising from the estimator variance $\text{Var}[\expect{A}_I^{Est}] = \sum_i b_i^2 \text{Var}[\expect{A}(a_i \epsilon)]$. Assuming the variance of each noisy estimate is equal, the total number of samples must be increased by a {\em multiplicative} factor of $\sum_i b_i^2$ in order to reduce the final estimator variance to that of the original estimates. Multiplicative factors ranging from $168$ (quadratic elimination) to $6.5E6$ (seventh order elimination) make variance reduction by additional sampling, which scales poorly as $N^{-1/2}$ where $N$ is the number of samples, impractical in light of device drift coupled with limited sampling per job submission in a cloud access model. Due to this rapid variance growth we simply utilize 1024 shots per noisy estimate and focus our attention on the performance and convergence of linear extrapolations. 

Note that the baseline $\theta_{10}$ is determined by comparing the phase difference between $\ket{1}$, as generated by a single $X_\pi$ pulse taken from the default pulse library, and the initial $\ket{0}$ state as illustrated by the solid horizontal lines in Fig.~\ref{fig:1} (b). We now compute the estimator error, $\Delta_{\theta_{01}} = |\expect{\theta_{01}}_I - \expect{\theta_{01}(\epsilon)}_E|$, as a function of the dataset size $D$. Qubit 0's shortest programs ($M=5$) show no discernible relaxation effects and linear extrapolation, with a slope $m\approx0$, converges near the $\Delta \theta \approx 11$ as seen in Fig.~\ref{fig:noise_factors_vs_rabi_amplitude_qb0} (b). Qubit 12's estimate is initially thrown off by the outlier, yet converges to within 10\% of the noiseless limit as the data-set size increases. With the exception of $M=160$ the linear extrapolations for higher values of $M$ converge to within $20,30\%$ for qubit 0 but to within $10-20\%$ for qubit 12 as seen in panel (b) in ~Figs.~\ref{fig:noise_factors_vs_rabi_amplitude_qb0},\ref{fig:noise_factors_vs_rabi_amplitude_qb12}. 

In contrast to high order extrapolations, using a linear extrapolation but increasing the number of samples, i.e. the data set size $D$, the variance can be substantially reduced. The linear extrapolations, utilizing multiple stretched logically equivalent programs of $M$ Rabi flops and the standard error, due to statistical sampling and $b_i$-weighting, are provided in the bottom panels of Figs.~\ref{fig:noise_factors_vs_rabi_amplitude_qb0},~\ref{fig:noise_factors_vs_rabi_amplitude_qb12}. Overall, we see that qubit 12 performs significantly better than qubit 0 in over the entire range of $M$-cycles over which linear extrapolation converges. These results shed light on the physical origins of single qubit extrapolation and should be considered when selecting the best of qubits for a given program and QEM methodology.

\section{Conclusion}

In this work we have explored the concept of benchmarking qubit performance with respect to an analog extrapolation-based error mitigation strategy. Our work shows how, in contrast with previous single qubit metrics such as fidelity, mitigation benchmarking measures the performance of qubits with respect to the noise properties that are central to the removal of the effects of quantum noise processes in post processing. This opens a future research avenue to a device-wide benchmarking protocol in order to determine the optimal set of qubits (including benchmarking the mitigability of two-qubit gates) on various qubit platforms. Future programs should be executed on spatially localized sets of qubits which conform with the domain algorithm and minimize the expexted $\Delta_A$ in order to maximize post-processed estimator accuracy. We also note that it will be interesting to correlate the results of such an analysis with fidelities and other device information provided by hardware vendors. 

\acknowledgments

E.\ F.\ D. and R.\ C.\ P.\, acknowledge DOE ASCR funding under the Quantum Testbed Pathfinder program, FWP number ERKJ332. J.\ W.\ O.\ G.\ was supported by the Department of Energy Science Undergraduate Laboratory Internship (SULI) program. 
This research used quantum computing system resources supported by the U.S. Department of Energy, Office of Science, Office of Advanced Scientific Computing Research program office. Oak Ridge National Laboratory manages access to the IBM Q System as part of the IBM Q Network. The views expressed are those of the authors and do not reflect the official policy or position of IBM or the IBM Q team.

%%%%%%%% Bibliography %%%%%%%%%%
\bibliography{refs}
\end{document}